\documentclass[iop]{emulateapj}

\shorttitle{A Successful Broad-band Survey for Giant \lya\ Nebulae II}
\shortauthors{Prescott et al.}

\newcommand{\bw}{$B_{W}$}
\newcommand{\ergss}{erg~s$^{-1}$}

\newcommand{\civ}{\hbox{{\rm C}\kern 0.1em{\sc iv}}}
\newcommand{\ciii}{\hbox{{\rm C}\kern 0.1em{\sc iii}}]}
\newcommand{\heii}{\hbox{{\rm He}\kern 0.1em{\sc ii}}}
\newcommand{\halpha}{\hbox{{\rm H}\kern 0.1em$\alpha$}}
\newcommand{\hbeta}{\hbox{{\rm H}\kern 0.1em$\beta$}}
\newcommand{\neiv}{\hbox{{\rm Ne}\kern 0.1em{\sc iv}}]}
\newcommand{\oii}{[\hbox{{\rm O}\kern 0.1em{\sc ii}}]}
\newcommand{\oiii}{[\hbox{{\rm O}\kern 0.1em{\sc iii}}]}
\newcommand{\lya}{Ly$\alpha$}

\newcommand{\numfirst}{39} 
\newcommand{\numsecond}{40} 
\newcommand{\numthird}{6} 
\newcommand{\numfirsttarg}{15} 
\newcommand{\numsecondtarg}{5} 
\newcommand{\numthirdtarg}{6} 
\newcommand{\searcharea}{8.5} 

\begin{document}

\title{A Successful Broad-band Survey for Giant \lya\ Nebulae II: Spectroscopic Confirmation} 

\author{Moire K. M. Prescott\altaffilmark{1,2}, Arjun Dey\altaffilmark{3}, Buell T. Jannuzi\altaffilmark{3}}

\altaffiltext{1}{TABASGO Postdoctoral Fellow; Department of Physics, Broida Hall, Mail Code 9530, University of California, Santa Barbara, CA 93106; mkpresco@physics.ucsb.edu}
\altaffiltext{2}{Steward Observatory, University of Arizona, 933 N. Cherry Avenue, Tucson, AZ 85721}
\altaffiltext{3}{National Optical Astronomy Observatory, 950 North Cherry Avenue, Tucson, AZ 85719} 

\begin{abstract}
Using a systematic broad-band search technique, we have carried out a survey for large \lya\ nebulae 
(or \lya\ ``blobs'') at $2\lesssim z\lesssim3$ within \searcharea\ square degrees of the NOAO 
Deep Wide-Field Survey (NDWFS) Bo\"otes field, corresponding to a total survey comoving volume 
of $\approx$10$^{8}$ h$_{70}^{-3}$ Mpc$^{3}$.  Here, we present our spectroscopic observations of candidate 
giant \lya\ nebulae.  Of 26 candidates targeted, 5 were confirmed to have \lya\ emission at 
$1.7\lesssim z \lesssim 2.7$, four of which were new discoveries. The confirmed \lya\ nebulae 
span a range of \lya\ equivalent widths, colors, sizes, and line ratios, and most show spatially-extended 
continuum emission.  The remaining candidates did not reveal any strong emission lines, but 
instead exhibit featureless, diffuse blue continuum spectra.  Their nature remains mysterious, but we 
speculate that some of these might be \lya\ nebulae lying within the redshift desert (i.e., $1.2 \lesssim z \lesssim 1.6$).  
Our spectroscopic follow-up confirms the power of using deep broad-band imaging to search for the bright 
end of the \lya\ nebula population across enormous comoving volumes. 
\end{abstract}


\section{Introduction}
\label{sec:intro}

Giant radio-quiet \lya\ nebulae (also know as ``\lya\ blobs'') 
provide a window into the physics of ongoing massive galaxy formation 
\citep[e.g.,][]{francis96,ivi98,stei00,pal04,mat04,dey05,sai06,nil06,smi07,pres09,yang09}.  
Rare systems found primarily in overdense regions \citep{pres08,mat04,mat05,mat11,sai06,yang09,yang10}, 
\lya\ nebulae are extremely luminous ($L_{Ly\alpha}\sim10^{44}$ erg s$^{-1}$)
and are frequently associated with young, star-forming galaxy populations 
or obscured AGN \citep[e.g.,][]{basu04,mat04,dey05,gea07,gea09,pres12b}.  
Theoretical and observational investigations into the power source behind 
\lya\ nebulae have painted a varied picture, presenting arguments 
for AGN powering \citep{chap04,basu04,gea07,gea09}, starburst-driven winds 
\citep{tani00,tani01,mat04}, spatially-extended star formation 
\citep{mat07,pres12b}, cold accretion 
\citep{nil06,smi08,goerdt10,faucher10,rosdahl12}, or some combination thereof \citep{dey05,pres09,pres12b,webb09,colbert11}.  
Larger samples of giant radio-quiet \lya\ nebulae drawn from unbiased surveys are needed in order 
to accurately measure the space density of these sources, to determine the emission 
mechanisms primarily responsible for the \lya\ nebula class as a whole, and to understand their 
relationship to the more well-studied \lya\ haloes found around many 
high redshift radio galaxies (e.g., \citealt{mcc93}, and references therein; 
\citealt{ojik96,wei05,miley06,barrio08,smith09,zirm09}).

Increasing the sample of such a rare class of objects requires surveying large 
comoving volumes, but the standard approach relying on narrow-band imaging 
is limited by observational expense.  Our complementary approach was to design 
a new search technique using deep {\it broad-band} imaging \citep{presphd}.  
We employed this method to select a sample of \lya\ nebula candidates from 
the Bo\"otes Field of the NOAO Deep Wide-Field Survey \citep[NDWFS;][]{jan99}.  
A full presentation of our survey algorithm as well as the selection function of this approach 
and the implications for the space density of \lya\ nebulae can be found in companion 
papers \citep[][2013, in preparation; hereafter, Papers I and III, respectively]{pres12a}.  

In this work (Paper~II), we focus on the spectroscopic follow-up of \lya\ 
nebula candidates drawn from deep archival broad-band imaging of the NDWFS Bo\"otes field.  
In Section~\ref{sec:design}, we give a brief outline of the survey design and candidate sample.  
In Section~\ref{sec:follow-up}, we discuss our spectroscopic observations and reductions.  
Section \ref{sec:indiv} presents the 5 \lya\ nebulae 
discovered within this \searcharea\ square degree survey and discusses our primary contaminant sources, 
potentially an interesting population in their own right.  Section \ref{sec:discussion} 
discusses the implications of this sample, and we conclude in Section \ref{sec:conclusions}.  

We assume the standard $\Lambda$CDM cosmology ($\Omega_{M}=0.3$, $\Omega_{\Lambda}=0.7$, $h=0.7$); 
1\arcsec\ corresponds to a physical scales of 8.3-7.8~kpc for redshifts of $z=1.2-2.9$.  All magnitudes 
are in the AB system \citep{oke74}.


\section{Search Design}
\label{sec:design}

We have designed an innovative search for \lya\ nebulae in deep broad-band imaging and applied it 
to the NDWFS Bo\"otes field (Paper~I).  
Our search is most sensitive to the largest and brightest \lya\ nebulae because it leverages 
the deep blue (\bw) imaging of NDWFS to look for sources where bright \lya\ emission 
boosts the broad-band flux relative to the very dark sky.  
Thanks to the wide area ($\approx$9~deg$^2$) of the NDWFS and the large width of the \bw\ filter 
($\approx1275$\AA, corresponding to $\Delta z \approx 1$), our survey is able to probe 
an enormous comoving volume ($\approx$10$^{8}$~h$_{70}^{-3}$ Mpc$^{3}$) with archival data and significantly 
reduce the required observational overhead.  Our survey is therefore complementary 
to smaller volume surveys for \lya\ nebulae that rely on sensitive narrow-band imaging 
\citep[e.g.,][]{mat04,sai06,smi07,yang09,yang10,mat11}.  

\begin{figure}
\center
\includegraphics[angle=0,width=6.5in]{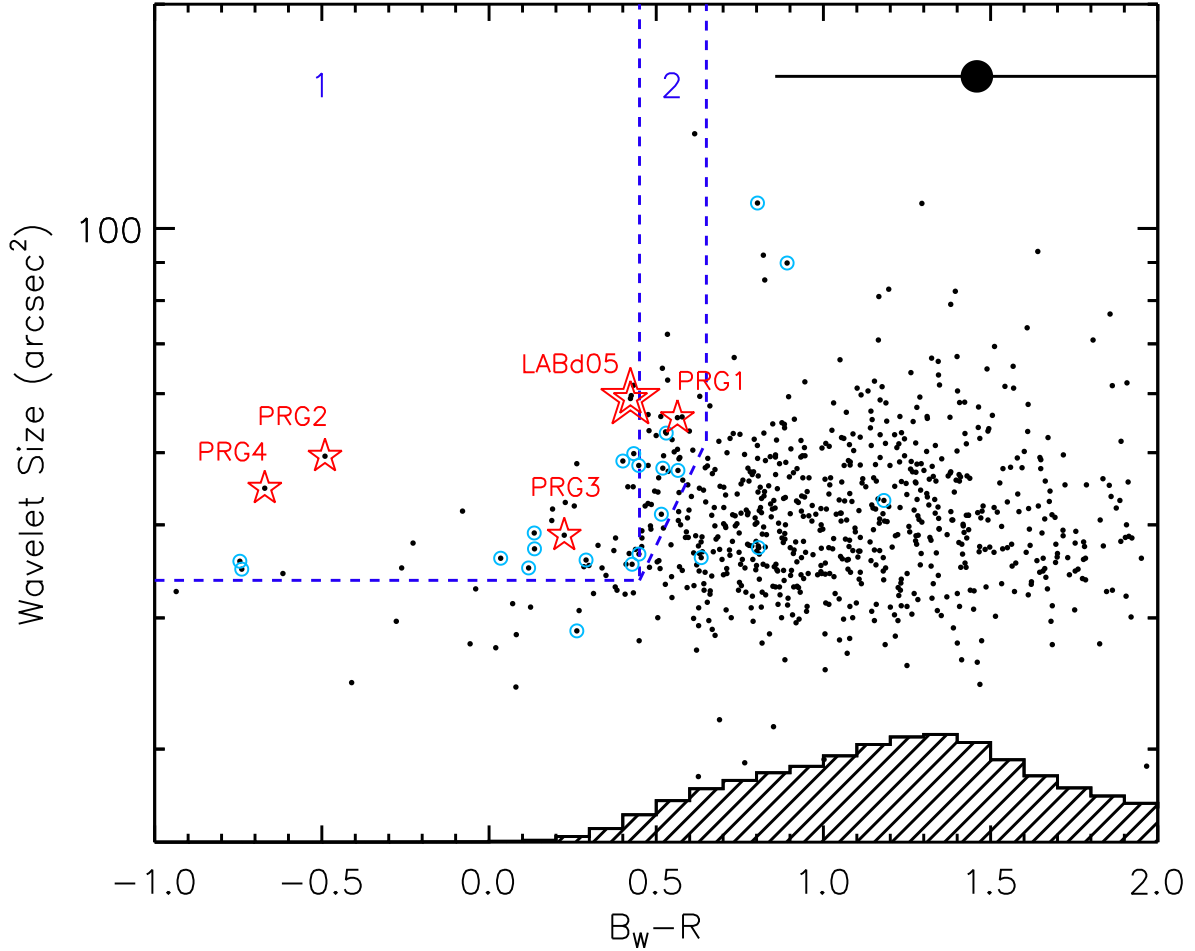} 
\caption[Size vs. $B_{W}-R$ of \lya\ Nebula Candidates]
{Size vs. $B_{W}-R$ color of \lya\ nebula candidates selected using the broad-band morphological search pipeline.
The wavelet size corresponds to the size of the source in the wavelet power map, but does not indicate 
the true nebular size of the object (Paper~I).  
All candidates are shown as filled black circles.  
The first and second priority selection regions indicated (blue dashed lines) contain \numfirst\ and \numsecond\ \lya\ 
nebula candidates, respectively.
The spectroscopic targets that showed \lya\ are indicated using red stars, 
and those that did not show \lya\ are marked with light blue circles.  The large filled
black circle with an error bar represents the typical color of low surface brightness galaxies \citep[LSBs;][]{hab07}, and the
histogram (plotted on a linear scale) represents the distribution of $B_{W}-R$ colors for field galaxies in NDWFS,
demonstrating that the colors of our final \lya\ nebula candidates are substantially bluer than typical LSBs and
field galaxies.
}
\label{fig:fullcandidate}
\end{figure}

The search algorithm and candidate sample are discussed in detail in Paper I.  
In brief, we selected giant \lya\ nebula candidates using wavelet analysis of the 
compact-source-subtracted \bw\ images. We selected a set of first and second priority candidates 
(Figure~\ref{fig:fullcandidate}) based on their $B_{W}-R$ color, as measured using large 
30~pix=7.7\arcsec\ diameter apertures, and wavelet size, as determined using 
SourceExtractor \citep{ber96} on the wavelet-deconvolved images.  
The final candidate sample consisted of \numfirst\ first priority and \numsecond\ second priority sources 
over a search area of \searcharea\ square degrees.  Both first and second priority samples contained 
candidates with diffuse morphologies (morphological category {\sc diffuse}) as well as those that appear to be tight 
groupings of compact sources (morphological category {\sc group}), as discussed in Paper~I.  
In addition, we flagged \numthird\ sources from outside these selection regions that showed 
promising morphologies (third priority).
 
\section{Spectroscopic Follow-up}
\label{sec:follow-up}

In this section we describe the spectroscopic follow-up of the \lya\ nebula candidate sample.

\subsection{Observations \& Reductions}

We targeted a total of 26 Lya nebula candidates (\numfirsttarg\ first priority, \numsecondtarg\ second priority, 
and all \numthirdtarg\ third priority candidates).  Longslit spectroscopic observations were obtained using 
the MMT and the Blue Channel Spectrograph during 2007 May, 2008 April and June (Table~\ref{tab:obs}).
The observations used the 300 l/mm grating with 1\farcs0/1\farcs5 wide slits resulting in a 
resolution FWHM of 2.6/3.4\AA\ and a wavelength range of $\Delta\lambda\approx 3100-8320$\AA. 

During our 2007 run, we chose slit orientations based primarily on the morphology of the candidate (i.e., 
aligned with the major axis of the emission as estimated from the \bw-band morphology), and as a secondary criterion 
attempted to intersect a nearby bright reference object if possible.  In practice, however, we found that 
the faintness of our candidates and the short duration of our spectroscopic exposures necessitated a positional 
reference.  As a result, slit orientations during the 2008 runs were chosen to always include a positional reference object, 
and consequently do not always trace the major axis of the \bw\ morphology.  During these later runs, we 
dithered the target along the slit by $\approx 5\arcsec$ between exposures to minimize the effect of any bad pixels.  
The full list of targeted candidates and the results of the spectroscopic observations are given 
in Table~\ref{tab:target}.  

We reduced the spectroscopic data using IRAF.\footnote{IRAF is distributed by 
the National Optical Astronomy Observatories, which are operated by the Association of Universities for 
Research in Astronomy, Inc., under cooperative agreement with the National Science Foundation.} 
We subtracted the overscan and bias, and applied a flat-field correction using normalized observations of 
the internal quartz flat-field lamps.  
Twilight flats were used to determine the illumination correction for the science frames.  
We removed cosmic rays from the 2D sky-subtracted data 
using {\it xzap}.\footnote{http://iraf.noao.edu/iraf/ftp/iraf/extern/xdimsum020627}  
One-dimensional spectra were generated using the task {\it apall} and optimal variance-weighted 
extraction \citep{val92}; the spectral trace was determined using bright unresolved sources on 
the slit.  
We determined the wavelength solution to an rms precision of $\approx$0.08-0.18\AA\ 
using HeArNe and HgCd comparison lamps, and then corrected the data for any slight systematic 
offset using the night sky lines as a reference.  The final wavelength calibration is 
accurate to $\pm$0.42\AA.  The relative flux calibration was based on 
observations of the standard stars BD+33 2642, BD+26 2606, BD+28 4211, Feige 34, 
and Wolf~1346.\footnote{KPNO IRS Standard Star Manual}  
For each night, we applied a grey shift to compensate for any variable grey 
(i.e., independent of wavelength) extinction that may have affected a given standard star observation 
relative to one taken under better conditions.
The sensitivity functions derived from individual standard star exposures 
were consistent to within $\lesssim$0.1~mag.  

Due to the faintness of the candidates and the fact that we are searching for luminous \lya\ nebulae at $z\approx2-3$, 
the aim of our follow-up spectroscopic program was to look for strong, high equivalent width line emission.  
A single strong line in the blue can be identified as \lya\ rather than an unresolved \oii$\lambda\lambda$3726,3729 doublet 
(the only other possibility at these wavelengths) 
due to the fact that a detection of \oii\ would be accompanied by stronger detections of \oiii$\lambda\lambda$4959,5007 and \halpha\ as well.  
Candidates with strong \lya\ emission at $z\approx2-3$ in the \bw-band are easily detectable with the MMT/Blue Channel 
down to a 5$\sigma$ limit of $\approx1-7\times10^{-17}$ in 30 minutes (assuming a FWHM$_{obs}=12$\AA\ \lya\ line 
measured within a 1\farcs5$\times$5\arcsec\ aperture).  This corresponds to limiting line 
luminosities of $\approx0.3-5\times10^{42}$~\ergss, which are below the typical luminosities of giant \lya\ nebulae.  
However, continuum-only sources are much fainter and require longer integration times to yield high 
signal-to-noise ratio spectra, more than was available during our spectroscopic campaign.  
To allow us to target the largest number of candidates, we therefore carried out a quick reduction of 
the data in real-time and continued integrating on each of the 26 targeted candidates up until the 
point where we could either confirm the presence of a line or confirm the presence of continuum with no strong lines.  
The deeper spectroscopy necessary for studying the spectral properties of our confirmed \lya\ nebulae in detail 
as well as for detecting absorption features in continuum-only sources is left to future observations with larger 
telescopes.

\section{Results}
\label{sec:indiv}

Of the \numfirsttarg\ first priority and \numsecondtarg\ second priority candidates targeted for spectroscopic follow-up, 
4 first priority sources and 1 second priority source had confirmed \lya\ emission: 
we easily recovered the previously-discovered large \lya\ nebula at 
$z\approx2.66$ \citep[LABd05;][]{dey05} and discovered new, spatially-extended \lya\ nebulae 
at $z\approx1.67$, $z\approx1.88$, $z\approx2.14$, and $z\approx2.27$.  
In addition, we also targeted \numthirdtarg\ third priority candidates that showed promising 
diffuse morphologies upon visual inspection. 
However, no \lya\ or \oii\ line emission was confirmed in any of the third priority candidates.  
In this section, we describe each of the confirmed sources in turn and then discuss the primary contaminants 
to our survey.  

\subsection{Confirmed \lya\ Sources}
\label{sec:confirmed}

\begin{figure}
\center
\includegraphics[angle=0,width=6.5in]{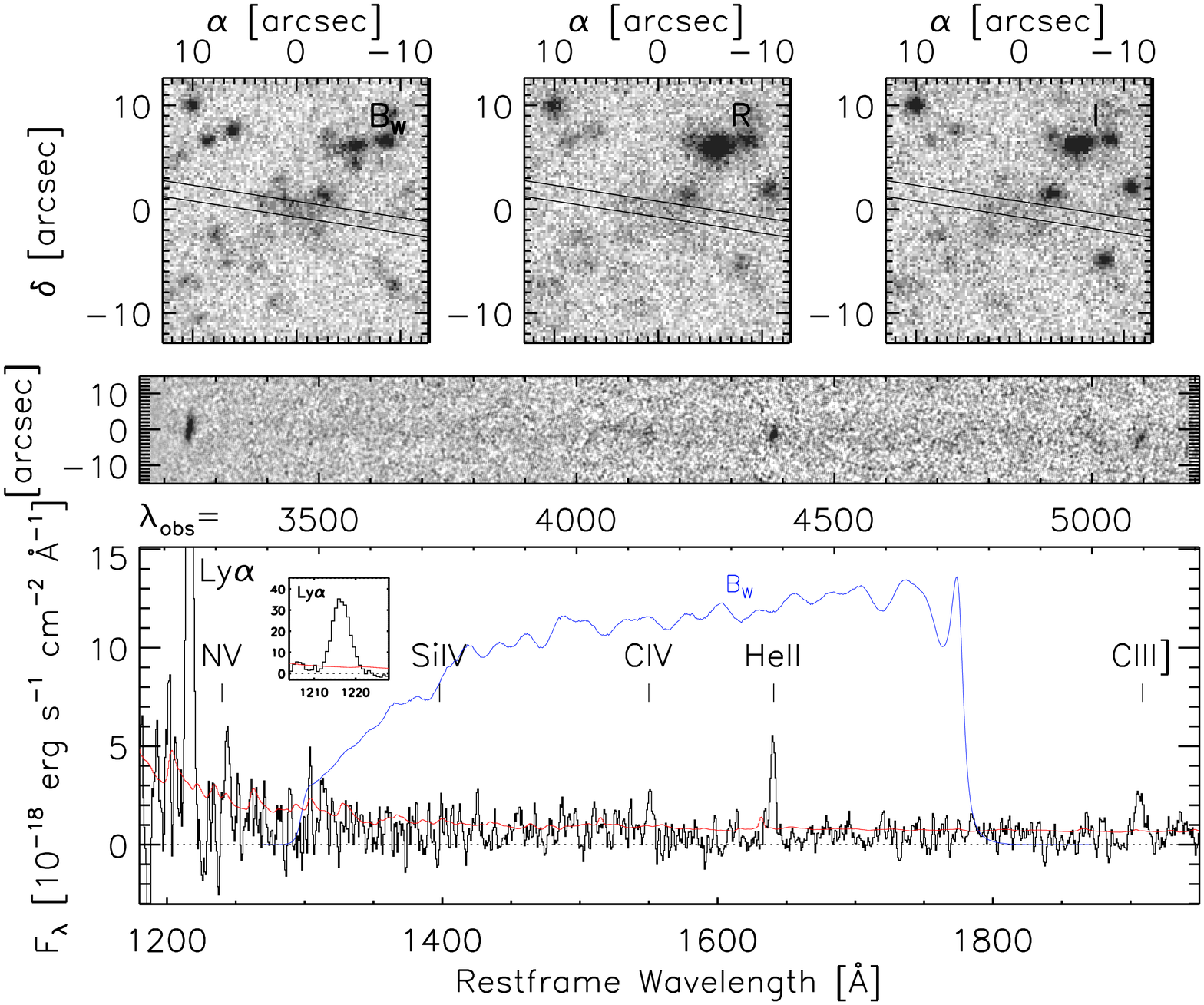} 
\caption[PRG1]
{Imaging and MMT spectroscopic observations of PRG1, a \lya\ nebula at $z\approx1.67$.  The 
top row shows the optical \bw, $R$, and $I$ imaging along with the slit used for 
follow-up spectroscopy.  The central panel contains the two-dimensional spectrum 
versus observed wavelength, and the bottom panel presents the one-dimensional spectrum 
extracted from a 1.5$\times$5.04\arcsec\ aperture.  The error spectrum (red line) and 
\bw\ bandpass (blue line) are shown, and the positions of common emission lines are 
indicated.  The one-dimensional \lya\ emission line profile is shown in the inset panel.  
Spectroscopic data presented are from the second night of observations 
only \citep[UT 2008 June 9; for details see][]{pres09}. 
}
\label{fig:prg1spec}
\end{figure}

\begin{figure}
\center
\includegraphics[angle=0,width=6.5in]{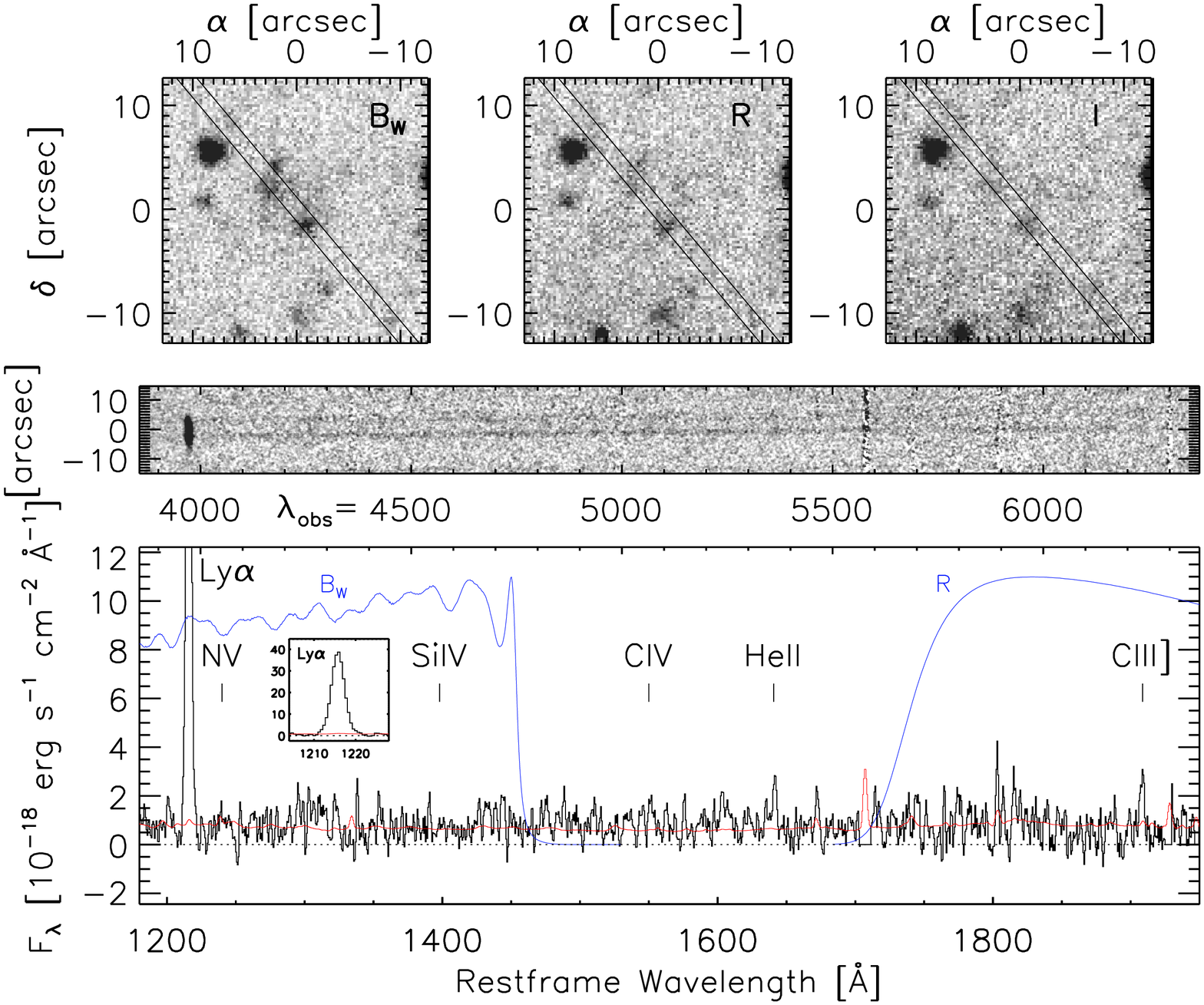} 
\caption[PRG2]
{Imaging and MMT spectroscopic observations of PRG2, a \lya\ nebula at $z\approx2.27$.  The 
top row shows the optical \bw, $R$, and $I$ imaging along with the slit used for 
follow-up spectroscopy.  The central panel contains the two-dimensional spectrum 
versus observed wavelength, and the bottom panel presents the one-dimensional spectrum 
extracted from a 1.5$\times$7.84\arcsec\ aperture.  The error spectrum (red line) and 
\bw\ and $R$ bandpasses (blue lines) are shown, and the positions of common emission lines are 
indicated.  The one-dimensional \lya\ emission line profile is shown in the small inset panel.}
\label{fig:diamondspec}
\end{figure}

\begin{figure}
\center
\includegraphics[angle=0,width=6.5in]{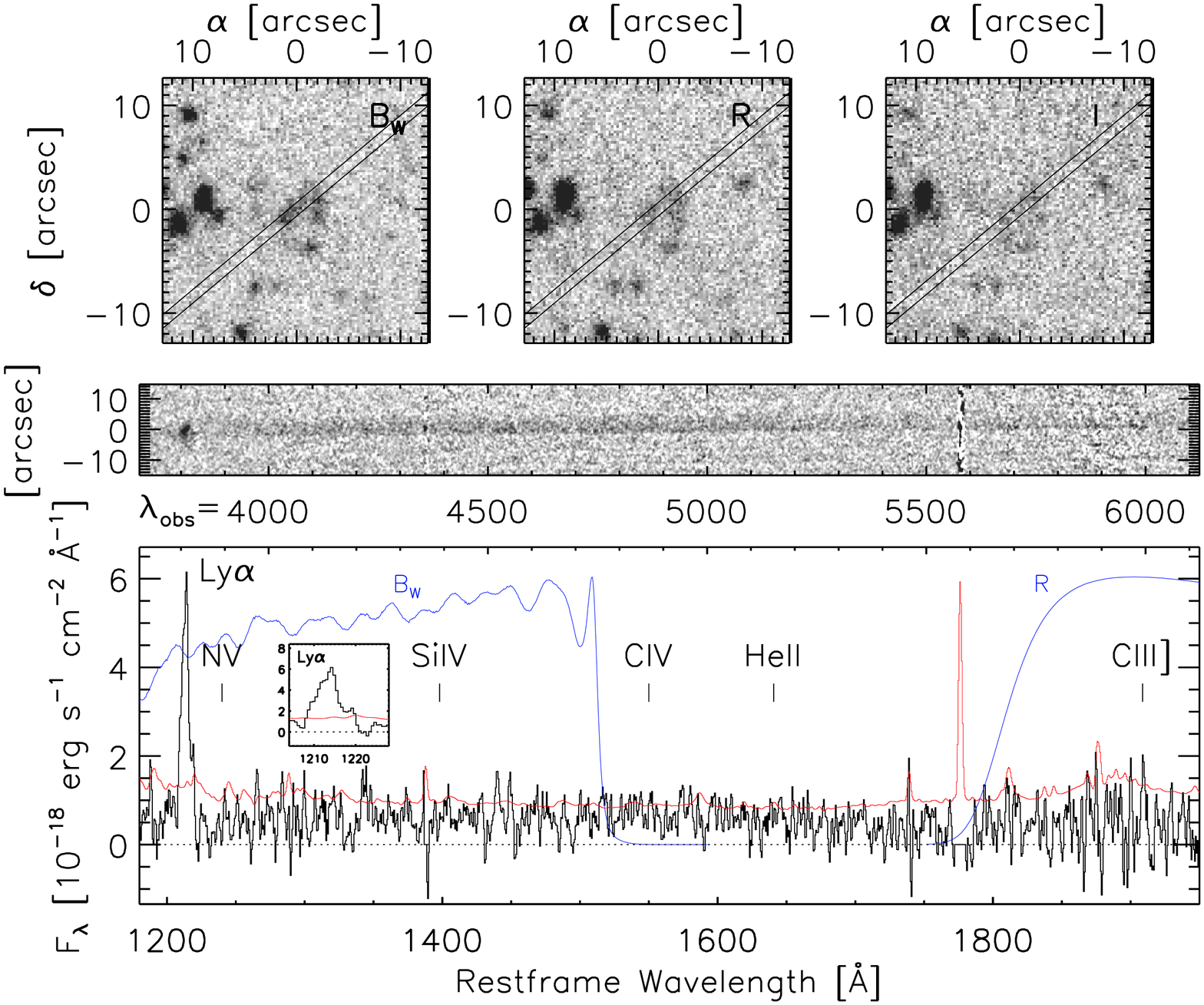} 
\caption[PRG3]
{Imaging and MMT spectroscopic observations of PRG3, a \lya\ nebula at $z\approx2.14$.  The 
top row shows the optical \bw, $R$, and $I$ imaging along with the slit used for 
follow-up spectroscopy.  The central panel contains the two-dimensional spectrum 
versus observed wavelength, and the bottom panel presents the one-dimensional spectrum 
extracted from a 1.0$\times$5.6\arcsec\ aperture.  The error spectrum (red line) and 
\bw\ and $R$ bandpasses (blue lines) are shown, and the positions of common emission lines are 
indicated.  The one-dimensional \lya\ emission line profile is shown in the small inset panel.}
\label{fig:horseshoespec}
\end{figure}

\begin{figure}
\center
\includegraphics[angle=0,width=6.5in]{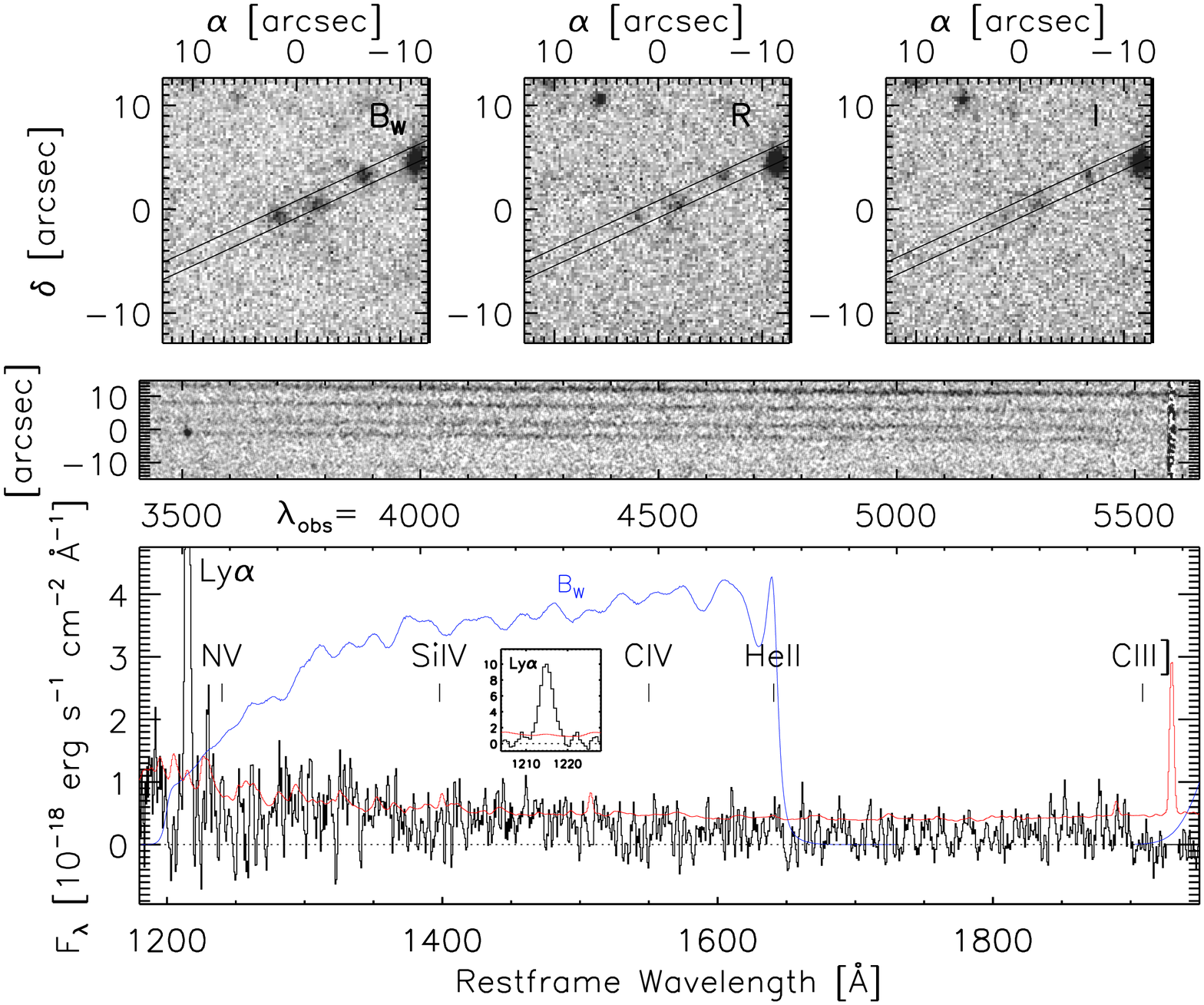} 
\caption[PRG4]
{Imaging and MMT spectroscopic observations of PRG4, a \lya\ nebula at $z\approx1.89$.  The 
top row shows the optical \bw, $R$, and $I$ imaging along with the slit used for 
follow-up spectroscopy.  The central panel contains the two-dimensional spectrum 
versus observed wavelength, and the bottom panel presents the one-dimensional spectrum 
extracted from a 1.5$\times$1.68\arcsec\ aperture.  The error spectrum (red line) and 
\bw\ and $R$ bandpasses (blue lines) are shown, and the positions of common emission lines are 
indicated.  The one-dimensional \lya\ emission line profile is shown in the small inset panel.  
The size of PRG4 is unknown; the source 
is very compact along the spectroscopic slit, but additional diffuse emission that 
may or may not be \lya\ is visible to the south in the \bw\ imaging.
}
\label{fig:tinyspec}
\end{figure}

\begin{figure}
\center
\includegraphics[angle=0,width=6.5in]{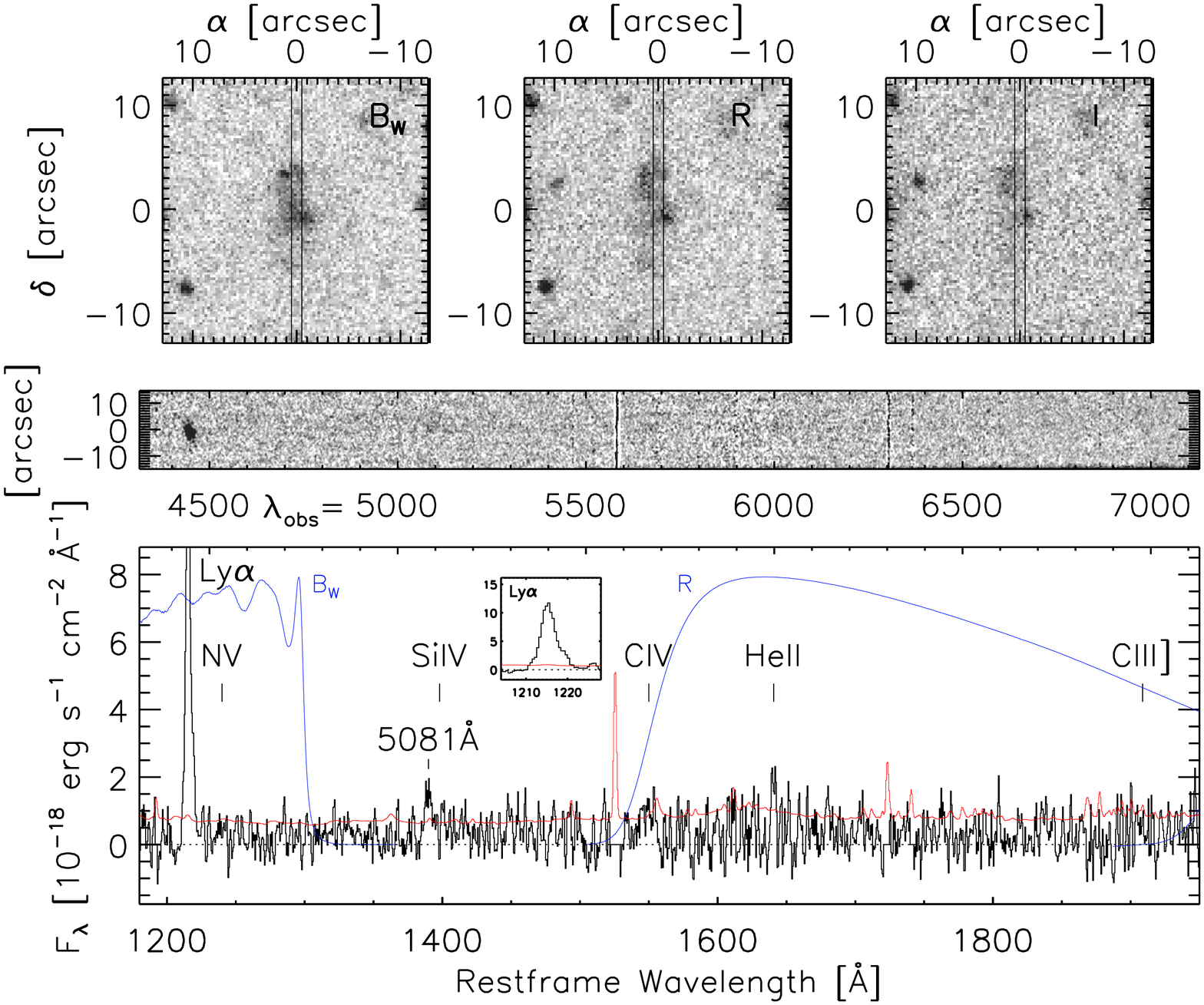} 
\caption[Dey et al. (2005) \lya\ Nebula]
{Imaging and MMT spectroscopic observations of LABd05, a previously-known 
\lya\ nebula at $z\approx2.656$ \citep{dey05}.  The 
top row shows the optical \bw, $R$, and $I$ imaging along with the slit used for 
follow-up spectroscopy.  The central panel contains the two-dimensional spectrum 
versus observed wavelength and the bottom panel presents the one-dimensional spectrum 
extracted from a 1.0$\times$4.48\arcsec\ aperture.  The error spectrum (red line) and 
\bw\ and $R$ bandpasses (blue lines) are shown, and the positions of common emission lines are 
indicated.  The one-dimensional \lya\ emission line profile is shown in the small inset panel.  
The emission line at 5081\AA\ from a background galaxy at $z\approx3.2$ 
is labeled for reference \citep{dey05}.
}
\label{fig:deyspec}
\end{figure}

Figures~\ref{fig:prg1spec}-\ref{fig:deyspec} show the postage stamps, two-dimensional spectra, 
and one-dimensional spectra of the \lya\ sources; the measured properties are listed 
in Table~\ref{tab:candidate}.  The spectral extraction apertures were chosen to maximize 
the signal-to-noise ratio of \lya.  Redshifts were determined from the centroid of 
a Gaussian fit to the observed \lya\ line.  No correction was included for \lya\ absorption, 
a potential source of bias in our redshift estimates.  
The \bw\ sizes, isophotal areas, and surface brightnesses were measured above the median 1$\sigma$ surface brightness limit 
of the entire NDWFS survey from the original \bw\ images using 
SourceExtractor \citep[{\sc detect\_minarea}$=$5, {\sc detect\_thresh}$=$28.9 mag arcsec$^{-2}$;][]{ber96}. 
The \lya\ sizes were measured from the 2D spectra using SourceExtractor above the 1$\sigma$ surface brightness 
limit at the location of \lya\ ({\sc detect\_minarea}$=$5, 
{\sc detect\_thresh}$\approx1\times10^{-18}$ erg s$^{-1}$ cm$^{-2}$ \AA$^{-1}$ arcsec$^{-2}$; see Table~\ref{tab:candidate}).  
The \bw\ sizes along the slit can underestimate the \lya\ sizes measured from the 2D spectra; in the case of PRG3, our deepest 
spectrum, the \bw\ size underestimates the \lya\ size by a factor of $\gtrsim$1.3. 

Estimates for the total \lya\ isophotal area and total \lya\ luminosity are also given in Table~\ref{tab:candidate}.  
The approximate total \lya\ isophotal area was estimated by correcting $A_{B_{W}}$ by a factor of $\nu^{2}$, where $A_{B_{W}}$ 
is the isophotal area of the source on the \bw\ image and $\nu$ is the ratio of the \lya\ and \bw\ sizes measured along the slit.  
The approximate total \lya\ luminosity was derived by scaling the \lya\ luminosity within the spectroscopic aperture 
by the geometric correction factor $f_{geo}=A_{B_{W}}\times\nu^{2}/(\omega \times d)$, 
where $\omega$ is the slit width and $d$ is the spatial extent of the spectral extraction aperture.  
We stress, however, that in using area corrections based on broad-band imaging we are relying on the 
assumptions that the \bw\ emission roughly traces the \lya\ emission in the source and that the relative 
factor between the \bw\ and \lya\ sizes is the same throughout the object as it is along the slit; 
either of these assumptions may be violated in practice.   

In what follows, we discuss each confirmed \lya\ source in detail.  
Of the confirmed \lya\ nebulae, four were originally categorized as having a {\sc diffuse} morphology 
while one (PRG4) was categorized as having a {\sc group} morphology (see Paper~I).

\subsubsection{PRG1}  
\label{sec:prg1}
PRG1 is a remarkable \lya\ nebula (Figure~\ref{fig:prg1spec}).  As discussed in \citet{pres09}, it is 
the first example of a \lya\ nebula with strong, spatially-extended \heii\ emission and weak metal lines, 
suggestive of a hard ionizing continuum and potentially low metallicity gas.  
The \bw\ imaging shows a diffuse nebula and several compact sources, the brightest of which is 
located at the northwest edge of the nebula.  Despite the strong \lya\ emission and large size 
($>$78~kpc), PRG1 was selected as a second priority candidate because, 
at $z\approx1.67$, \lya\ is at the edge of the optical window and not 
contained within the \bw\ band, giving the source a relatively red $B_{W}-R$ color.  
Thanks to its diffuse blue continuum (95\%) and \heii\ emission (5\%), however, this source was still 
selected by our survey.  When first discovered, this source was the lowest redshift \lya\ nebula known 
and the only one that shows strong spatially-extended \heii\ emission; PRG1 is therefore an ideal target 
for detailed study of the physical conditions and kinematics within \lya\ nebulae.  
Analysis of the metallicity and source of ionization in PRG1 is given in \citet{pres09}, 
and more detailed analysis using deep Keck/LRIS spectroscopy is in progress (Prescott et al., in preparation).

\subsubsection{PRG2}  
PRG2 is a large \lya\ nebula at $z\approx2.27$ with a roughly diamond-shaped morphology 
in the \bw\ image (Figure~\ref{fig:diamondspec}).  
The identification of the strong line in the spectrum as \lya\ is secure based on the fact that no 
other lines are well-detected in the discovery spectrum and corroborated by weak detections at the 
positions of \civ$\lambda\lambda$1548,1550, \heii$\lambda$1640, and \ciii$\lambda$1909.  In the case of \oii\ at lower redshift, we would have easily 
detected \oiii\ and \halpha\ instead.  
The \lya\ nebula spans almost 100~kpc, and at the southwestern corner there is a very blue compact 
source that appears to be a \lya-emitting galaxy from the spectrum.  
The redshift of this source is ideal for follow-up NIR spectroscopy as the rest-frame optical 
emission lines (\oii, \oiii, \hbeta, and \halpha) will be observable in the $J$, $H$, and $K$ bands. 
Continuum emission is observed from two compact knots located at either end and 
from spatially-extended emission at fainter levels in between them.  

\subsubsection{PRG3}  
PRG3 is a \lya\ nebula at $z\approx2.14$ (Figure~\ref{fig:horseshoespec}).  
It has a rather clumpy horseshoe-shaped morphology in the \bw\ imaging and spans 
$\approx$74~kpc.  The single strong line is identified as \lya\ rather than \oii\ based on the 
fact that we do not see corresponding detections of \oiii\ and \halpha.  The spectrum shows spatially-extended 
continuum, but no other strong emission lines.  

\subsubsection{PRG4} 
PRG4 appears to be a candidate that was selected due to a close grouping of compact blue sources 
(Figure~\ref{fig:tinyspec}).  Due to the very blue color, it was flagged as a high priority target.  
At these wavelengths (blueward of the restframe wavelength of \oii), \lya\ is the only possible strong line.  
In addition, no other strong lines are seen in the spectrum.  
Although the \bw\ size of the full grouping is roughly 7\arcsec, the observed \lya\ 
at $z\approx1.89$ is only marginally extended along the direction of the spectroscopic slit (3.9\arcsec, 33~kpc).  
The source may be larger in \lya: there is additional diffuse emission outside the slit that is visible to the southwest 
in the \bw\ imaging, but without further spectroscopy, we cannot determine if it is associated with coincident \lya\ emission.  

\subsubsection{LABd05}  LABd05 is the source that was the inspiration for our broad-band \lya\ nebula 
search \citep[Figure~\ref{fig:deyspec};][]{dey05}.  
One of the largest \lya\ nebulae known \citep[$\gtrsim100$~kpc;][]{dey05}, it is located at $z\approx2.656$.  
Our shallow MMT spectrum was taken at a slightly different position than the existing 
deeper spectroscopy from Keck but shows a hint of \heii\ emission and an emission 
line at 5081\AA, both seen previously in the system \citep{dey05}.  
The emission line at 5081\AA\ is thought to be \lya\ from a background interloper galaxy 
at $z\approx3.2$, the compact source that is visible in the ground-based imaging and 
located at the western edge of the slit for this observation.  
Detailed study of ground-based data as well as high resolution imaging from HST showed that there 
are numerous compact galaxies, including a spectroscopically-confirmed Lyman break galaxy, within the system 
that are offset spatially from the \lya\ nebula itself \citep{dey05,pres12b}.  
The HST imaging demonstrated that the nebula contains diffuse restframe UV continuum emission, that the \lya\ 
emission itself is smooth with a relatively round and disk-like morphology, and that the \heii\ emission 
is spatially extended by $\approx0.6-1$\arcsec\ \citep[$\approx5-8$~kpc;][]{pres12b}.

\subsection{Survey Contaminants}
\label{sec:contam}

\begin{figure}
\center
\includegraphics[angle=0,width=7in]{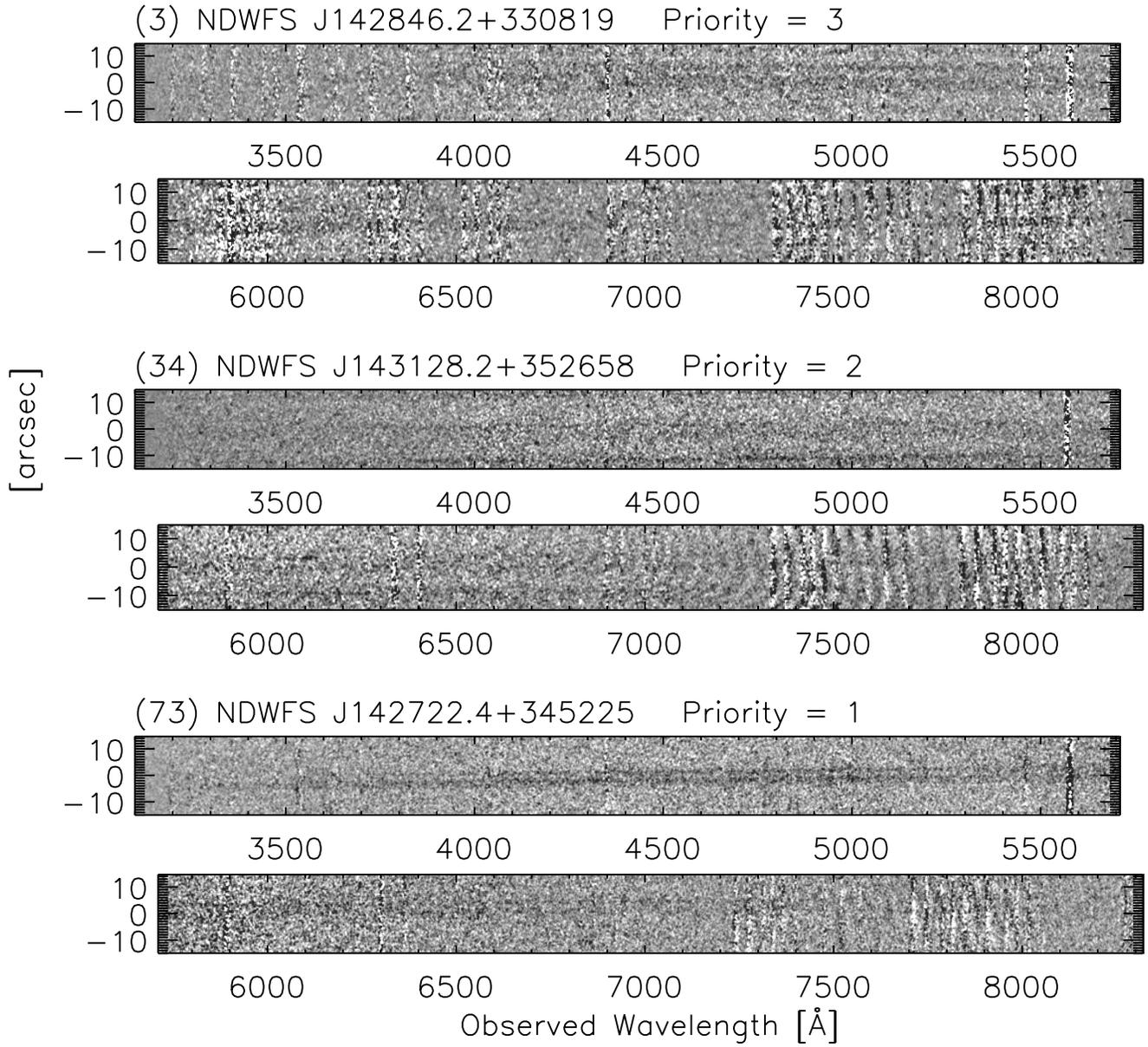} 
\caption[Blue continuum object]
{Examples of the continuum-only contaminant sources discussed in the text (Section~\ref{sec:contam}).  
Two-dimensional MMT/Blue Channel Spectrograph spectra are shown versus observed wavelength, with 
the target source centered at 0\arcsec\ along the spatial dimension. 
}
\label{fig:continuumonly}
\end{figure}

The dominant contaminants in both the first and second priority spectroscopic samples are 
sources with spatially resolved blue continuum emission but no visible emission lines.  
Despite the lack of strong line emission in the \bw\ band, our morphological broad-band 
search selected these sources either due to sufficiently extended, blue continuum emission 
or due to a close projected grouping of blue galaxies.  A few examples of these continuum-only 
sources are shown in Figure~\ref{fig:continuumonly}.  

Without deeper spectroscopy, we can only speculate as to the nature of these continuum-only sources.  
The largest cases within the candidate sample (sources 1+2 and 3; Paper~I) are so spatially extended 
($\approx$15-86\arcsec, which at $z\approx1.2-2.9$ would imply physical sizes of $\approx130-710$~kpc 
in the continuum) and irregular in morphology that they are almost certainly located within the Galaxy, 
perhaps low surface brightness Galactic reflection nebulae.  
Since low redshift ($z\lesssim1.2$) blue star-forming populations or low surface brightness galaxies 
(LSBs) would be expected to show \oii, \oiii, or \halpha\ emission lines in our spectra, some fraction of the remaining 
continuum-only contaminants may in fact be galaxies or \lya\ nebulae in the redshift desert ($1.2\lesssim z\lesssim1.6$), 
for which \lya\ is blueward of the atmospheric cut-off ($\lambda_{obs}\lesssim3100$\AA) 
but for which \oii\ has been redshifted past the red end of our MMT/Blue Channel spectra 
($\lambda_{obs}\gtrsim8320$\AA).  One of the \lya\ nebulae confirmed by our survey (PRG1, 
at $z\approx1.67$) is in fact below the redshift where \lya\ is covered by the \bw-band.  
Instead, this source was selected by our survey primarily due to blue continuum emission, and 
it was only thanks to the excellent blue sensitivity of MMT/Blue Channel that we were still able 
to detect the \lya\ emission at $\approx$3250\AA.  
The case of PRG1 lends credence to the hypothesis that at least a fraction of the continuum-only 
``contaminant" sources are in fact \lya\ nebulae at $1.2\lesssim z \lesssim1.6$.  

At the same time, however, our expectation from Paper~I was that \lya\ nebulae in the redshift desert 
should make up roughly 25\% of the candidate sample, under the optimistic assumption that the \lya\ nebula 
number density does not evolve significantly with redshift.  In practice, we found continuum-only 
detections represented a much larger fraction (75\%) of the target spectroscopic sample, 
suggesting that this explanation may not be the full story.  
While the presence of continuum emission in the spectra does confirm that these continuum-only sources 
are indeed real astrophysical objects and not artifacts within the NDWFS imaging, deeper ground-based 
optical spectroscopy or UV spectroscopy from space will be required to confirm their origin on a case 
by case basis.  At this stage, their nature remains mysterious.

\section{Discussion}
\label{sec:discussion}

\subsection{A Successful Broad-band \lya\ Nebula Survey}

This work is the first demonstration of the feasibility of conducting systematic 
surveys for large \lya\ nebulae using deep broad-band imaging datasets.  
The primary advantage of our unusual survey approach is the enormous comoving volume 
that can be surveyed using deep archival datasets.  
In addition, since this search technique is best used in the blue where the sky is dark, 
the resulting \lya\ nebula sample is weighted to lower redshifts ($z<3$) where we 
have the opportunity to undertake detailed studies of their properties.  
The obvious trade-off is that our approach is not as sensitive to \lya\ nebulae that are intrinsically faint, 
low surface-brightness, or compact in morphology, as discussed in Papers~I and III.   
Our search, therefore, provides a measurement of the bright end 
of the \lya\ nebula luminosity function, nicely complementing standard narrow-band surveys 
that probe to fainter luminosities.

The success rate for finding sources with \lya\ emission was 
$\approx27$\% for first priority and $\approx20$\% for second priority candidates.  
Therefore, if we were able to target all the \lya\ nebula candidates in our sample, 
we would expect to find a total of $\approx18$ \lya\ nebulae ($\approx10$ and $\approx8$ 
from the first and second priority set, respectively).  
While one of the goals of our broad-band survey for \lya\ nebulae 
is to place constraints on the space density of these rare objects, 
a robust estimate of the space density requires a detailed analysis of the 
selection function and is beyond the scope of the present paper.  
Here, we briefly discuss our detection rate in the context of traditional narrowband 
\lya\ nebula surveys. 

Based on the results of the narrowband survey carried out at $z\approx2.3$ 
by \citet{yang09}, in Paper~I we estimated the expected number of \lya\ nebulae 
in our survey volume to be $\sim$60-400, assuming a 100\% detection rate, the same 
detection limit as \citet{yang09}, and a constant volume density as a function of redshift.  
Instead, we have confirmed 5 \lya\ nebulae, and scaling these results to the unobserved 
candidates, we expect to find only 18.  While this estimate is extremely crude, it does 
suggest that the space density of the detected \lya\ nebulae in our sample is lower than 
that of the \citet{yang09} sample.  Possible reasons for this difference are that (a) the \citet{yang09} 
narrow-band survey is more sensitive to fainter, and therefore less luminous, \lya\ nebulae than our 
broad-band survey; (b) the \citet{yang09} survey does not exclude \lya\ nebulae with bright 
central sources whereas our survey does due to the nature of the morphological search algorithm; 
and/or (c) the \citet{yang09} survey is more sensitive to cosmic variance than our larger volume survey.  
We defer a more detailed discussion of the space density of \lya\ nebulae implied by our survey to Paper III.


\subsection{Dispersion within the \lya\ Nebula Class}

The power of a systematic survey is the opportunity it provides to find out what is common among a class 
of objects and also what the dispersion in properties is among members of that class.
The four large cases in our sample (PRG1, PRG2, PRG3, LABd05) span nearly an order of magnitude 
in total \lya\ luminosity ($50-170\times10^{42}$ erg s$^{-1}$), show a range of \lya\ equivalent 
widths ($\sim50-260$\AA), and are at least $70-100$~kpc in diameter.  
Morphologically, the four large \lya\ nebulae all show clumps and knots of emission in the broad-band imaging.  
The brightest compact knot in PRG1 is very red while that in PRG2 is remarkably blue.  
In addition, all four show what appears to be diffuse continuum emission in the ground-based spectroscopy.  
This could either be due to many unresolved clumps, or due to a continuum component that is truly spatially extended.  
Analysis of HST imaging of one system (LABd05) lends support to the latter hypothesis, 
revealing that most of the continuum in this one source is unresolved even at high resolution 
\citep[0.1\arcsec;][]{pres12b}.  In three cases (PRG1, PRG2, LABd05) there is evidence for emission in other lines 
(e.g., \civ, \heii, or \ciii).  

\begin{figure}
\center
\includegraphics[angle=0,width=5in]{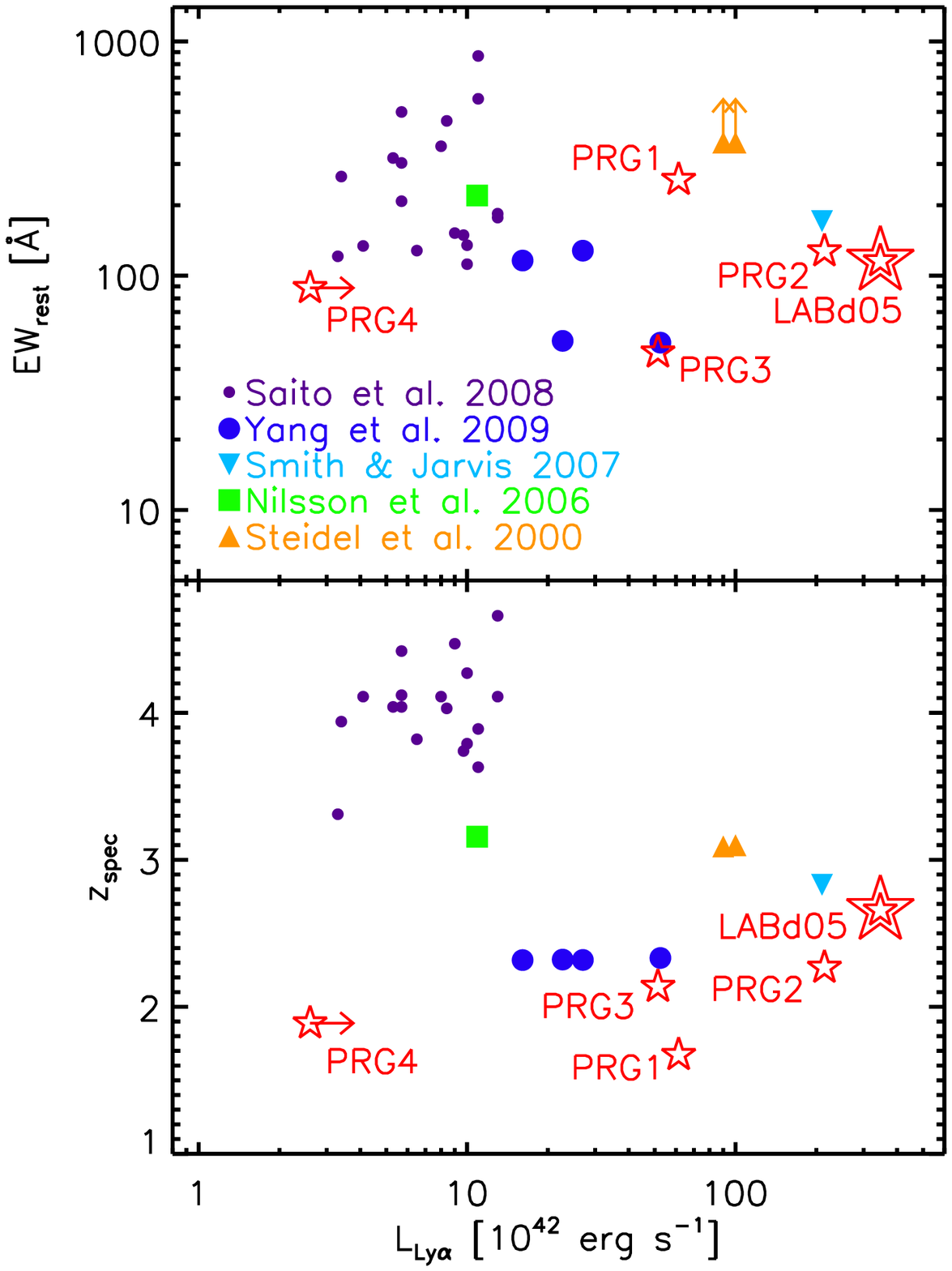} 
\caption[EW_{rest} versus \lya\ luminosity for \lya\ Nebulae]
{Rest-frame equivalent width (EW$_{rest}$, top panel) and spectroscopic redshift ($z_{spec}$, bottom panel) 
versus total \lya\ luminosity ($L_{Ly\alpha}$) 
for \lya\ nebulae reported in the literature \citep{stei00,nil06,smi07,sai08,yang09} 
and those found using our broad-band search (red stars).  
Our survey succeeded in discovering \lya\ nebulae with 
rest-frame equivalent widths comparable to those found by previous narrow-band surveys, 
but covers a large redshift range using just a single broad-band filter.
}
\label{fig:alllab}
\end{figure}

Given that diffuse continuum emission will have a larger impact on the observed broad-band color than 
line emission, one might ask if our survey is biased towards finding lower equivalent width 
sources than narrow-band surveys.  
In fact, however, Figure~\ref{fig:alllab} shows that our survey uncovered \lya\ nebulae with 
rest-frame equivalent widths comparable to those of luminous \lya\ nebulae found using standard 
narrow-band surveys but over a much larger redshift range.


\section{Conclusions}
\label{sec:conclusions}
We have carried out an innovative and economical systematic search for large \lya\ nebulae using archival 
deep, broad-band data.  While our technique is only sensitive to the largest and brightest \lya\ nebulae,  
it is able to probe enormous comoving volumes ($\approx10^{8}$ h$_{70}^{-3}$ Mpc$^{3}$) using existing deep 
broad-band datasets.  
The details of our search algorithm, the selection function, and implied space density 
are discussed in Papers~I and III of this series.  
In this paper (Paper~II), we presented details of our spectroscopic follow-up of \lya\ nebula candidates.  
Within our $\sim$\searcharea\ square degree survey area and a redshift range 
of $z\approx1.6-2.9$, we confirmed 4 new \lya\ nebulae and recovered 1 previously known case.  
The brightest 4 \lya\ nebulae have \lya\ luminosities of $\sim5-17\times10^{43}$ erg s$^{-1}$ and 
sizes of $>$70~kpc.  Our broad-band search found \lya\ nebulae with large \lya\ luminosities and 
equivalent widths comparable to those found with narrow-band surveys, but revealed a new 
common theme: at least some large \lya\ nebulae show diffuse, spatially-extended continuum emission.  
The primary contaminants in our survey are sources that show nothing but blue continuum in the optical 
range, some of which we suspect may be galaxies or \lya\ nebulae located in the redshift desert.  
Deep continuum spectroscopy and comparisons to {\it GALEX} photometry will be required to confirm this claim.  
This work uncovered the first example of a giant \lya\ nebula at $z<2$ and has demonstrated the feasibility 
of using deep broad-band datasets to efficiently locate luminous \lya\ nebulae within enormous comoving volumes.

\acknowledgments
We are grateful to Christy Tremonti and Kristian Finlator for observing assistance 
and to the telescope operators at the MMT, in particular Ale Milone and John McAfee.  
We would also like to thank the Steward Observatory TAC for the generous allocations of 
MMT time used in support of this project, as well as the anonymous referee for helpful suggestions.  
This research draws upon data from the NOAO Deep Wide-Field Survey (NDWFS) as distributed by the NOAO Science Archive.
NOAO is operated by the Association of Universities for Research in Astronomy (AURA), Inc. under
a cooperative agreement with the National Science Foundation.  
M. P. was supported by a NSF Graduate Research Fellowship and a TABASGO Prize Postdoctoral 
Fellowship.  A. D. and B. T. J.'s research is supported by NOAO, which is operated by 
AURA under a cooperative agreement with the National Science Foundation.


\begin{deluxetable}{cccccc}
\tabletypesize{\scriptsize}
\tablecaption{Observing Log}
\tablewidth{0pt}
\tablehead{
\colhead{UT Date} & \colhead{Instrumental Resolution\tablenotemark{a}} & \colhead{Unvignetted Slit} & \colhead{Spatial Binning} & \colhead{Seeing} & \colhead{Conditions}  \\
 & (arcsec) & (\AA) & (arcsec/pixel) & (arcsec) & \\
 }
\startdata

2007 May 20 & 2.6 & 1.0$\times$120 & 0.56 & 1.0-1.2 & Clear, high winds \\
2007 May 21 & 2.6 & 1.0$\times$120 & 0.56 & 1.0-1.2 & Clear, high winds \\
2007 May 22 & 3.4 & 1.5$\times$120 & 0.56 & 1.3-1.7 & Clear, high winds \\

2008 April 3  & 3.4 & 1.5$\times$120 & 0.28 & 1.0 & Mostly clear\\
2008 April 30 & 3.4 & 1.5$\times$120 & 0.28 & 1.2-1.9 & Clear, high winds \\

2008 June 8 & 3.4 & 1.5$\times$120  & 0.28 & 1.0 & Clear \\
2008 June 9 & 3.4 & 1.5$\times$120  & 0.28 & 1.1-2.0 & Clear, high winds \\

\enddata

\tablenotetext{a}{Quoted instrumental resolution is the average of measurements of the Hg~\textsc{i}$\lambda$4047, Hg~\textsc{i}$\lambda$4358, Hg~\textsc{i}$\lambda$5461, and O~\textsc{i}$\lambda$5577 sky lines.}
\label{tab:obs}
\end{deluxetable}

\begin{deluxetable}{cccccccccc}
\tabletypesize{\scriptsize}
\tablecaption{Spectroscopic Targets}
\tablewidth{0pt}
\tablehead{
  & \colhead{Candidate Name} & \colhead{Right Ascension} & \colhead{Declination} & \colhead{Priority} & \colhead{UT Date} & \colhead{Exposure} & \colhead{Class\tablenotemark{a}} & \colhead{Notes} \\
 & & (hours) & (degrees) & & & Time (s) & &  \\
 }
\startdata
    (2) &                   NDWFS J143006.9+353437 & 14:30:06.864 & 35:34:36.73 &    3 &       2007 May 20-21 &       2400 &       Continuum &       Galactic? \\ 
    (3) &                   NDWFS J142846.2+330819 & 14:28:46.228 & 33:08:19.42 &    3 &       2007 May 20-21 &      10800 &       Continuum &       Galactic? \\ 
   (10) &                   NDWFS J143411.0+331731 & 14:34:10.975 & 33:17:31.26 &    1 &          2007 May 20 &       1800 &           \lya\ &          LABd05 \\ 
   (14) &                   NDWFS J143512.3+351109 & 14:35:12.336 & 35:11:08.62 &    2 &         2008 Jun 8-9 &       7200 &           \lya\ &            PRG1 \\ 
   (18) &                   NDWFS J143222.8+324943 & 14:32:22.768 & 32:49:42.67 &    2 &           2008 Jun 8 &       1800 &       Continuum &               - \\ 
   (24) &                   NDWFS J142614.7+344434 & 14:26:14.714 & 34:44:34.22 &    1 &           2008 Jun 8 &       3600 &       Continuum &               - \\ 
   (26) &                   NDWFS J142622.9+351422 & 14:26:22.905 & 35:14:22.02 &    1 &           2008 Apr 3 &       3600 &           \lya\ &            PRG2 \\ 
   (29) &                   NDWFS J142526.3+335112 & 14:25:26.332 & 33:51:12.16 &    1 &           2008 Jun 9 &       3600 &       Continuum &               - \\ 
   (31) &                   NDWFS J142547.1+334454 & 14:25:47.126 & 33:44:54.13 &    1 &          2008 Apr 30 &       2400 &       Continuum &               - \\ 
   (33) &                   NDWFS J142714.8+343155 & 14:27:14.791 & 34:31:54.55 &    2 &           2008 Jun 8 &       1800 &       Continuum &               - \\ 
   (34) &                   NDWFS J143128.2+352658 & 14:31:28.245 & 35:26:57.91 &    2 &          2007 May 22 &       5400 &       Continuum &               - \\ 
   (40) &                   NDWFS J142653.2+343855 & 14:26:53.172 & 34:38:55.39 &    1 &          2008 Apr 30 &       3600 &           \lya\ &            PRG4 \\ 
   (44) &                   NDWFS J142927.8+345906 & 14:29:27.837 & 34:59:06.14 &    3 &          2007 May 21 &       3600 &       Continuum &               - \\ 
   (52) &                   NDWFS J143706.6+335653 & 14:37:06.588 & 33:56:52.65 &    2 &       2007 May 20-21 &       4800 &       Continuum &               - \\ 
   (58) &                   NDWFS J142516.6+324335 & 14:25:16.629 & 32:43:35.47 &    1 &           2008 Jun 8 &       3600 &       Continuum &               - \\ 
   (59) &                   NDWFS J143412.7+332939 & 14:34:12.722 & 33:29:39.19 &    1 &          2008 May 20 &      10800 &           \lya\ &            PRG3 \\ 
   (65) &                   NDWFS J143207.2+343101 & 14:32:07.224 & 34:31:01.34 &    3 &          2007 May 22 &       3600 &       Continuum &               - \\ 
   (66) &                   NDWFS J142539.9+344959 & 14:25:39.859 & 34:49:59.19 &    1 &           2008 Jun 8 &       3600 &       Continuum &               - \\ 
   (70) &                   NDWFS J142753.8+341204 & 14:27:53.762 & 34:12:04.10 &    1 &       2007 May 20-21 &       8400 &       Continuum &               - \\ 
   (71) &                   NDWFS J142600.8+350252 & 14:26:00.842 & 35:02:52.36 &    3 &           2008 Apr 3 &       3600 &       Continuum &               - \\ 
   (72) &                   NDWFS J142643.9+340937 & 14:26:43.850 & 34:09:36.82 &    1 &           2008 Jun 9 &       3600 &       Continuum &               - \\ 
   (73) &                   NDWFS J142722.4+345225 & 14:27:22.408 & 34:52:24.74 &    1 &           2008 Apr 3 &       3600 &       Continuum &               - \\ 
   (74) &                   NDWFS J142620.0+340427 & 14:26:19.982 & 34:04:27.01 &    1 &          2008 Apr 30 &       2400 &       Continuum &               - \\ 
   (80) &                   NDWFS J142548.3+322957 & 14:25:48.283 & 32:29:56.58 &    1 &           2008 Jun 9 &       3600 &       Continuum &               - \\ 
   (82) &                   NDWFS J142449.8+324743 & 14:24:49.761 & 32:47:42.61 &    1 &          2008 Apr 30 &       3600 &       Continuum &               - \\ 
   (85) &                   NDWFS J142533.0+343912 & 14:25:32.966 & 34:39:11.95 &    3 &           2008 Apr 3 &       3600 &       Continuum &               - \\ 
\enddata
\tablenotetext{}{Candidate ID numbers in the first column are the same as in Paper~I.}
\tablenotetext{a}{Spectroscopic targets were classified as either showing ``\lya" or ``Continuum" emission.}
\label{tab:target}
\end{deluxetable}

\begin{deluxetable}{cccccc}
\tabletypesize{\scriptsize}
\tablecaption{\lya\ Nebula Measurements}
\tablewidth{0pt}
\tablehead{
 & \colhead{PRG1} & \colhead{PRG2} & \colhead{PRG3} & \colhead{PRG4} & \colhead{LABd05} \\
 }
\startdata
                              Aperture (arcsec) & 1.5  $\times$ 5.04  & 1.5  $\times$ 7.84  & 1.0  $\times$ 5.60  & 1.5  $\times$ 1.68  & 1.0  $\times$ 4.48  \\
       $\lambda_{Ly\alpha,obs}$ (\AA) &  3249.61 $\pm$    0.39  &  3971.41 $\pm$    0.13  &  3813.28 $\pm$    0.90  &  3511.23 $\pm$    0.67  &  4444.99 $\pm$    0.31  \\
                             Redshift & 1.6731 $\pm$0.0003  & 2.2668 $\pm$0.0001  & 2.1368 $\pm$0.0007  & 1.8883 $\pm$0.0005  & 2.6564 $\pm$0.0003  \\
                           $F_{Ly\alpha}$ (10$^{-17}$ erg s$^{-1}$ cm$^{-2}$) &  44.1 $\pm$  4.0  &  49.2 $\pm$  1.1  &  10.2 $\pm$  1.2  &  10.3 $\pm$  1.2  &  19.0 $\pm$  0.9  \\
                                      $L_{Ly\alpha}$ (10$^{42}$ erg s$^{-1}$) &   8.2 $\pm$  0.7  &  19.3 $\pm$  0.4  &   3.5 $\pm$  0.4  &   2.6 $\pm$  0.3  &  10.9 $\pm$  0.5  \\
                                  \lya\ EW$_{rest}$ (\AA) & 257.1 $\pm$ 29.1  & 127.3 $\pm$  6.3  &  47.1 $\pm$  6.4  &  88.7 $\pm$ 11.8  & 115.5 $\pm$  9.5  \\
                                 \lya\ FWHM$_{obs}$ (\AA) &   9.19 $\pm$  0.60  &   8.52 $\pm$  0.19  &  23.36 $\pm$  7.90  &   6.51 $\pm$  0.89  &  15.44 $\pm$  0.70  \\
                         \lya\ $\sigma_{v}$ (km s$^{-1}$) &  361.2 $\pm$  23.7  &  273.9 $\pm$   6.1  &  782.1 $\pm$ 264.3  &  236.8 $\pm$  32.2  &  443.3 $\pm$  20.0  \\
               $F_{C\textsc{iv}\lambda1550}$ (10$^{-17}$ erg s$^{-1}$ cm$^{-2}$)\tablenotemark{a} &   2.1 $\pm$  1.1    &   1.8 $\pm$  0.8 & $<$  0.8 & $<$  0.8 & $<$  5.4  \\
               $F_{He\textsc{ii}\lambda1640}$ (10$^{-17}$ erg s$^{-1}$ cm$^{-2}$)\tablenotemark{a} &  5.8 $\pm$  1.0    &   1.8 $\pm$  0.9 & $<$  0.9    &   0.7 $\pm$  0.5    &   1.4 $\pm$  1.3  \\
               $F_{C\textsc{iii}\lambda1909}$ (10$^{-17}$ erg s$^{-1}$ cm$^{-2}$)\tablenotemark{a} &  4.8 $\pm$  0.9    &   3.0 $\pm$  1.1 & $<$  1.5 & $<$  1.4 & $<$  1.8  \\
 & & & & &  \\
\hline
 & & & & &  \\
       \bw\ Diameter Along Slit\tablenotemark{b} (arcsec) &  8.96  & 10.12  &  6.72  &  6.75  &  9.32  \\
      \bw\ Isophotal Area\tablenotemark{b} (arcsec$^{2}$) & 40.9  & 73.2  & 45.3  & 38.7  & 54.4  \\
       \bw\ Surface Brightness\tablenotemark{b} (mag arcsec$^{-2}$) & 27.2  & 27.0  & 26.8  & 27.0  & 27.0  \\
      \lya\ Diameter Along Slit\tablenotemark{c} (arcsec) &  9.24  & 12.04  &  8.96  &  3.92  &  8.96  \\
         \lya\ Diameter Along Slit\tablenotemark{c} (kpc) &  78.3  &  99.0  &  74.4  &  33.0  &  71.3  \\
             Approximate \lya\ Isophotal Area\tablenotemark{d} (arcsec$^{2}$) &  43.4  & 103.7  &  80.4  & $>$5.9\tablenotemark{f} &  50.3  \\
   Approximate Total $L_{Ly\alpha}$\tablenotemark{e} (10$^{42}$ erg s$^{-1}$) &  47.2 $\pm$  4.3  & 170.2 $\pm$  3.7  &  49.6 $\pm$  6.1  & $>$2.6\tablenotemark{f} & 122.8 $\pm$  6.0  \\
\enddata
\tablenotetext{a}{Line flux upper limits are 2$\sigma$ values.}
\tablenotetext{b}{\bw\ sizes, isophotal areas, and surface brightnesses measured from the NDWFS imaging using SourceExtractor with a detection threshold of 28.9 mag arcsec$^{-2}$, the median 1$\sigma$ \bw\ surface brightness limit of NDWFS.}
\tablenotetext{c}{\lya\ sizes measured from the 2D spectra using SourceExtractor with detection thresholds of [2.5, 1.0, 0.8, 2.1, 1.5]$\times10^{-18}$ erg s$^{-1}$ cm$^{-2}$ \AA$^{-1}$ arcsec$^{-2}$, the respective 1$\sigma$ line surface brightness limits at the position of \lya.}
\tablenotetext{d}{Approximate \lya\ isophotoal area computed by correcting the \bw\ isophotal area $A_{B_{W}}$ by a factor of $\nu^2$, where $\nu$ is the ratio of the \lya\ and \bw\ diameters measured along the slit.}
\tablenotetext{e}{Approximate total \lya\ luminosity computed by scaling the \lya\ luminosity measured within the spectroscopic aperture by a geometric correction factor of $f_{geo}=A_{B_{W}}\times\nu^{2}/(\omega \times d)$, where $A_{B_{W}}$ is the isophotal area of the source on the \bw\ image, $\nu$ is the ratio of the \lya\ and \bw\ diameters measured along the slit, $\omega$ is the slit width, and $d$ is the spatial extent of the spectral extraction aperture.}
\tablenotetext{f}{As the \bw\ emission is not an accurate tracer of the \lya\ emission in PRG4, approximate luminosity and area estimates have been replaced with lower limits derived from the spectroscopic data alone.}
\label{tab:candidate}
\end{deluxetable}

\end{document}